# Hands-Free VR

Jorge Askur Vazquez Fernandez*[1,2], Jae Joong Lee*[1], Santiago Andrés Serrano Vacca[2],
Alejandra Magana[3], Radim Peša[4], Bedrich Benes[1], and Voicu Popescu[1]

[1]*Department of Computer Science, Purdue University, West Lafayette, USA*
[2]*School of Engineering and Science, Tecnológico de Monterrey, Monterrey, Mexico*
[3]*Department of Computer and Information Technology, Purdue University, West Lafayette, USA*
[4]*Department of Informatics and Computers, Ostravska Universita, Ostrava, Czech Republic*

Abstract: We introduce Hands-Free VR, a voice-based natural-language interface for VR that allows interaction without additional hardware just using voice. The user voice command is converted into text using a fine-tuned speech-to-text deep-learning model. Then, the text is mapped to an executable VR command using an LLM, which is robust to natural language diversity. Hands-Free VR was evaluated in a within-subjects study (N = 22) where participants arranged objects using either a conventional VR interface or Hands-Free VR. The results confirm that Hands-Free VR is: (1) significantly more efficient than conventional VR interfaces in task completion time and user motion metrics; (2) highly rated for ease of use, intuitiveness, ergonomics, reliability, and desirability; (3) robust to English accents (20 participants were non-native speakers) and phonetic similarity, accurately transcribing 96.7% of voice commands, and (3) robust to natural language diversity, mapping 97.83% of transcriptions to executable commands.

## 1 INTRODUCTION

Virtual reality (VR) provides users powerful immersive visualizations of complex virtual environments (VEs). One of VR's great strengths is its natural interface for specifying the desired view based on tracking the user's head position and orientation. However, conventional VR interfaces are not equally effective when allowing users to configure, search, or modify the VE. Canonical tasks such as object creation, search, and selection, which are building blocks of complex interactions, often require repeated and tedious attempts to invoke, activate, tune, and undo operations through interface constructs that can be unfamiliar and unintuitive to the user. These challenges compound when applying the same command to multiple objects, such as isolating objects of a certain type or arranging objects in a specific configuration.

Conventional VR interfaces often adapt non-immersive controls to virtual environments, resulting in inefficiencies. For example, the familiar mouse becomes a 3D laser pointer that must be aimed precisely in mid-air without the haptic feedback provided by the physical stability of a desk, making fine object selection difficult. Additionally, while modern VR headsets feature built-in tracking that removes location constraints, navigating large virtual spaces in smaller physical areas remains challenging. It can require complex, disorienting solutions like redirection.

Although VR can replicate non-immersive controls-such as offering a 2D display + mouse interface or a deeply nested floating menu—these options break immersion and hinder the workflow. VR-specific interfaces (Section 2) improve efficiency but lose the familiarity of traditional controls.

Recent large language models (LLMs) handle complex language so effectively that their users can say what they want, and the LLM translates their words into executable code. This voice-to-code approach reduces inefficiency and frustration in VR controls, and as LLM-based code generation improves, these benefits become increasingly accessible.

This paper introduces Hands-Free VR, a voice-based VR interface. The user issues a natural-language voice command, a speech-to-text model for diverse English accents, and converts to text. An LLM then maps it to a unique executable VR command. Both models run on a workstation connected wirelessly to the headset. Compared to conventional VR controls that force users to walk around to select and arrange objects repeatedly, Hands-Free VR enables isolating and arranging them in fewer steps via

---

*These authors contributed equally to this work.

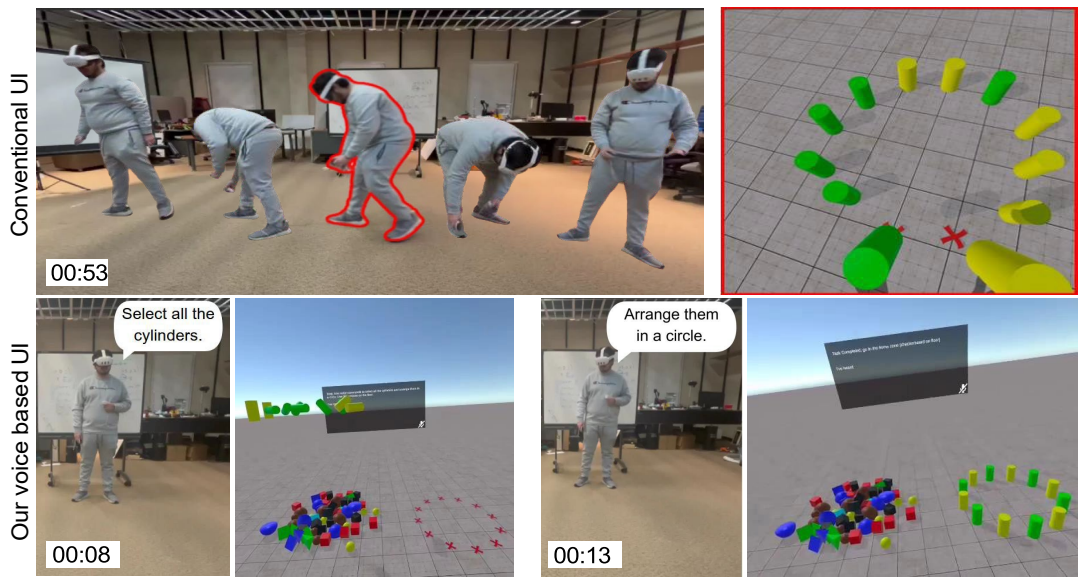

Figure 1: Conventional interface versus our Hands-Free VR. The task is to find all cylinders in a pile of objects and place them in a circle. With the conventional interface, the user has to walk to grab each cylinder and place it, needing 53s to complete the task. With our voice-based interface, the user first selects all the cylinders and then places them in a circle, using natural language spoken commands, completing the task in 13s.

voice commands (see Figure 1 and the video).

We evaluated Hands-Free VR in a controlled within-subjects study with 22 participants, approved by our Institutional Review Board (IRB). Participants completed object-finding and arrangement tasks using either a conventional VR interface (grab, carry, drop) or Hands-Free VR. With 20 participants being non-native English speakers, Hands-Free VR still achieved high accuracy: it correctly transcribed voice commands 96.71% of the time and converted them into executable VR commands 97.83% of the time, demonstrating robustness to accents, phonetic similarities, and natural language diversity.

We do not advocate using voice commands exclusively in VR. In applications where physical action is essential, voice should not bypass core interactions. However, we do believe voice can free users from tedious, repetitive, and non-essential tasks that hinder the application's main purpose. We claim the following contributions:

1. Hands-Free VR is a voice-based VR interface that is efficient and robust to diverse accents, phonetic similarities, and language variations.
2. A robust speech-to-text deep learning to accents.
3. A Large Language Model with Retrieval Augmented Generation for custom VR commands.

## 2 RELATED WORK

We review prior work on conventional VR interfaces that do not rely on the user's voice and voice-based VR user interfaces.

**Conventional VR User Interfaces:** Interactions in VR have unique challenges for many user interface tasks [Mine, 1995], and we limit the discussion to selection and text entry: tasks relevant to our study.

*Selection* is a challenging VR interaction task that has been extensively studied [Argelaguet and Andujar, 2013]. Users select targets using a virtual ray [Andujar and Argelaguet, 2007, Steinicke et al., 2006] or directly with their hands [Ware, 1990, Han and Wan, 2010]. A key challenge is the 3D visibility discrepancy between the user's eyes and hand [Argelaguet et al., 2008], where some visible targets may be unreachable from a natural hand position. In cluttered scenes, selection volumes instead of rays [Forsberg et al., 1996, Pierce et al., 1997] help by reducing the precision required to select small or occluded targets. Our voice-based interface complements VR selection methods, allowing users to select objects based on known features, regardless of size or occlusion.

*Text entry* is a notoriously difficult problem in VR, as typing on air keyboards is slow, inaccurate, and tiring. Innovative solutions have been proposed to address this issue, such as the implementation of virtual QWERTY keyboard layouts controlled using finger and thumb gestures [Fashimpaur et al., 2020], or such as the attachment of customized keycaps to the VR headset [Hutama et al., 2021]. Our deep learning speech-to-text solution offers intuitive operation and robustness to various accents and has the potential to support text entry by dictation.

**Voice-Based VR User Interfaces:** *Voice assistants* enhance VR applications like retail [Morotti et al., 2020], offering greater efficiency and user preference over traditional GUIs [Buchta et al., 2022c, Buchta et al., 2022b]. Our work uses LLM advancements to increase user freedom and interface robustness.

*Navigation* in VR benefits from voice interfaces, enabling users to teleport to distant virtual locations without physical movement [Hombeck et al., 2023, Calandra et al., 2022]. Natural Language Understanding [Zhao et al., 2020] simplifies interaction by removing the need for complex commands. Our work builds on this, fine-tuning a state-of-the-art speech-to-text model [Radford et al., 2023] with 13 English accents and demonstrating our voice-based interface for selection, posing, and navigation tasks.

*Conversation* highlights the power of voice-based interfaces, as reviewed in [Go¨bl. et al., 2021]. Recent LLMs like GPT-3 [Brown et al., 2020], PaLM [Chowdhery et al., 2022], LLaMA [Touvron et al., 2023], and ChatGPT [OpenAI, 2022] enable robust language understanding. Our work utilizes LLMs to map user commands to VR actions, focusing on robustness over conversation, and generates diverse language variants of VR commands offline.

*Object manipulation* in VR via voice has been introduced in CAD [Chu et al., 1997] and distant object interaction [Whitlock et al., 2018]. Our approach removes the need to memorize commands, allowing users to describe intentions in their words using an LLM, though it does not enable instant interpretation.

The *interface* is crucial to the user's VR experience. Voice interfaces free the user's hands for tasks [Monteiro et al., 2021] and are often preferred over GUIs, which feel tedious [Buchta et al., 2022a]. Hands-Free VR enables intuitive interaction without memorizing or practicing predefined commands.

## 3 OVERVIEW

The pipeline of Hands-Free VR is given in Figure 2. The user speaks a command, the VR headset captures their voice, and an edge server receives the audio data wirelessly and converts them to text with a robust deep learning model in diverse English accents and phonetic similarity to words. Next, a large language model (LLM) maps the transcribed text to executable VR commands using Retrieval-Augmented Generation (RAG) [Lewis et al., 2021] as shown in Section 4. The command is sent back to the VR headset and then is executed in the application.

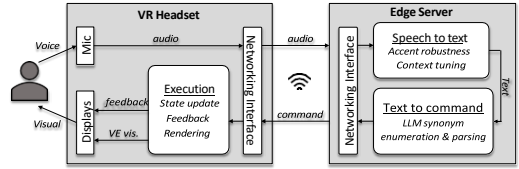

Figure 2: Hands-Free VR pipeline.

## 4 ROBUST SPEECH TO COMMAND CONVERSION

Understanding user intentions is important for a robust voice-based VR interface. This requires accurate handling of accents, phonetic similarities, and mapping free-form text into executable commands. Here, we detail our approach to ensuring speech-to-text and text-to-command robustness.

### 4.1 Data Synthesis for Robustness Enhancement

In the first offline step, our approach synthesizes data to support the robustness of speech-to-text and text-to-command. The data synthesis pipeline (Figure 3) proceeds in reverse order compared to the run-time order: data synthesis starts from the syntax and lexicon (1) and generates a set of natural language and a set of accent-diversified audio files (7).

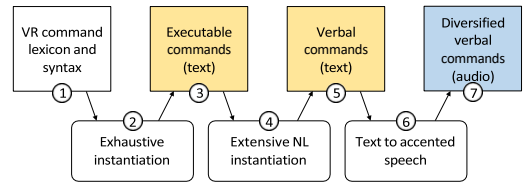

Figure 3: Text (yellow) and audio (blue) synthesis data to support robust speech-to-text and text-to-command conversion. The data is used as shown in Fig. 4.

The VR application's syntax and lexicon are used to generate all possible executable commands by assigning values to syntactic elements like verbs, objects, and attributes (e.g., *select(Cube)*, *select(Pyramid, yellow)*, *arrange(row)*). Each command is input into ChatGPT [OpenAI, 2022], which generates tens of natural language variants. These verbal commands are converted to speech using Amazon Polly [Amazon, 2016], producing tens of thousands of audio files with diverse expressions and 13 English accents.

The selection, relocation, and arrangement tasks of our user study (Section 6) are covered by a lexicon with 5 verbs, 5 objects, and 12 attributes (9 colors and 3 types of alignment), and by a syntax with sentences with just a verb, with a verb and an object, and with a verb, an object, and an attribute. Syntax and lexicon instantiation produced 66 executable commands. The 66 executable commands seeded 2,253 natural language variants, for an average of 34.14±10.52 variants per executable command. The minimum, median, and maximum variants per executable command are 4, 40, and 66, respectively. The 2,253 variants were converted from text to accented speech, resulting in 29,237 audio files. We refer the reader to the supplemental material FineTuningData.zip, which maps executable commands to natural language variants and audio files.

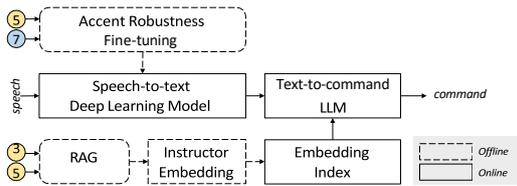

Figure 4: Speech-to-text and text-to-command robustness through fine-tuning and through Retrieval Augmented Generation (RAG) using the command text and audio data generated as shown in Fig. 3.

### 4.2 Robustness Enhancement

The synthesized data is used to improve (a) the speech-to-text and (b) the text-to-command conversion as shown in Figure 4.

(a) The English command in text form (input 5 in Figure 3) and the diversified English commands in audio form (7 in Figure 3) are used to fine-tune the Whisper [Radford et al., 2023] speech-to-text deep learning model using 5,000 iterations on our audio files, instead of training from scratch to utilize pre-trained dataset knowledge. (b) The (English command, executable command) pairs are used for Retrieval Augmented Generation (RAG) [Lewis et al., 2021] with a pre-trained LLM [Chung et al., 2022] and embedding [Su et al., 2023]. The embedding space is trained on instruction tasks, so we put our data in an instruction-like text format to use the embedding space [Su et al., 2023] as "`special_command of {English command} is {executable command}`".

### 4.3 Speech-to-Command Conversion

At run-time, the user gives a verbal command such as "select all red boxes". The command is converted to a text command by the speech-to-text deep learning model, with robustness to the user's spoken English accent and to words with similar sounds, e.g., preferring "boxes" to "foxes". The text command is interpreted by the LLM, which locates within its latent space the executable VR command that best aligns with the intended action. The LLM is queried with the text-based prompt "`what is the _special_command of {text command}?`". In this example, the LLM replies with the executable command, i.e., "`select(cube, red)`". Finally, the VR command is executed.

## 5 IMPLEMENTATION

The edge server (Figure 2) has an Intel i9-12900k 4.8GHz CPU, 128GB RAM, and NVIDIA RTX3090 GPU using Python 3.9 and PyTorch 2.0.1. The client is on a Meta Quest 3 [Meta, 2013] VR headset. The VR application is in Unity 3D [Li et al., 2018], version 2021.3.8f1. The server and client were connected with a 6E Wi-Fi.

The *speech-to-text* model is fine-tuned based on Whisper [Radford et al., 2023] with 29,237 audio files covering 13 different English accents. We use a mixed-precision floating point, AdamW [Loshchilov and Hutter, 2017], using a batch size 32 with a $10^{-5}$ learning rate. The well-known speech-to-text model metric, the word error rate, is used as our evaluation.

The Hands-Free VR *text-to-command* model combines LLM [Chung et al., 2022] with Instructor [Su et al., 2023] embedding space, directing multiple words (English language command) to one word (executable command), by utilizing RAG [Lewis et al., 2021] for its cost-effectiveness and customization. We formulate data for prompt augmentation by combining possible commands from users and their corresponding executable commands as "`special_command of {text} is {command}`" with a vector database [chromadb, 2022]. `special_command` is a unique identifier for our specific task to avoid any overlap with data from the original training dataset. Once the user's voice command was converted to text, we queried the user's question to the embedding as "`What is the special_command of {text}?`", and we retrieved from the LLM the most relevant answer, i.e., the executable command, using LangChain [LangChain, 2022]. We apply 4-bit quantization [Tim Dettmers, 2022] to the LLM to fit the 24GB VRAM GPU.

# 6 USER STUDY

We conducted an IRB-approved user study comparing Hands-Free VR to a conventional VR interface.

## 6.1 Methods

**Participants:** We have recruited N = 22 participants from our university. The study included 13 participants aged 18-25, 8 aged 26-30, and 1 over 30. Four identified as women, 17 as men, and 1 chose an alternative option. No participants identified as "Beginner" in English; 4 were "Intermediate," 16 "Advanced," and 2 "Native Speakers." One listed English as their native language, 5 Mandarin, 4 Hindi, and 12 chose "Other".

**Study design:** We opted for a within-subjects study design, with each participant performing the tasks in each condition. The design brings the advantage of greater statistical power for fewer participants. The learning effects are minor, as the two interfaces are substantially different from each other. The N = 22 participants are sufficient to detect effects of a large/very large size (i.e., Cohen's $d = 1.0$) with 0.90 power at a significance level $\alpha = 0.05$.

**Tasks:** Participants performed two tasks: In Task 1 (T1), they moved a subset of 96 virtual objects (spheres, hemispheres, pyramids, cubes, and cylinders) from a pile to a nearby box (1 m away). Objects were 10 cm tall, randomly colored, and the box measured 100 cm × 100 cm × 100 cm. In Task 2 (T2), they arranged objects from the pile into a row, matrix, or circular pattern, guided by red crosses on the floor, 2 m from the pile.

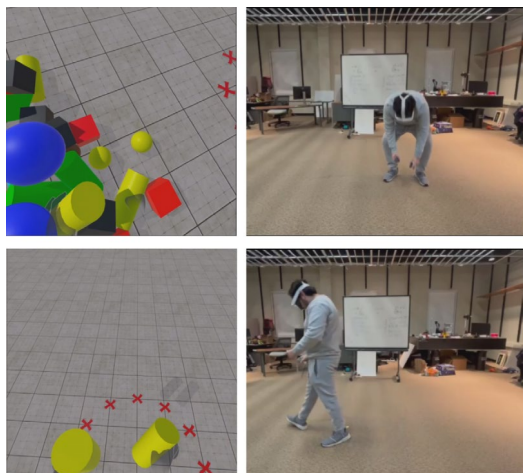

Figure 5: Control condition, task 2: The left column shows the participant's VR view as they collect two yellow cylinders (top) and place them in a circular pattern marked by red crosses (bottom).

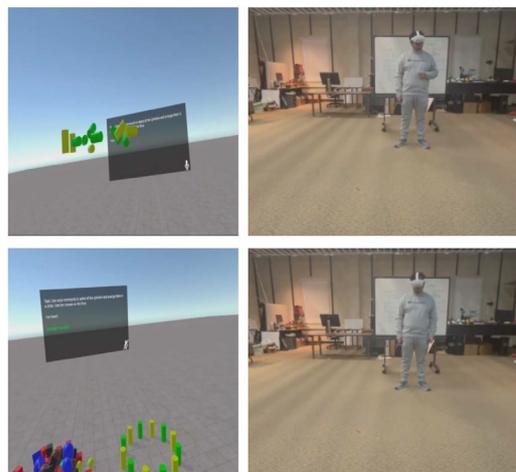

Figure 6: Experimental condition, task 2. The user says "Grab all the cylinders", which are lifted from the pile of objects (top), and then the user says "Put them in the circle" for arranging the cylinders on the floor (bottom).

**Conditions:** In the control condition (CC), participants used a conventional VR interface to manipulate objects with handheld controllers (Figure 5). They grabbed objects by pressing and holding the trigger button, releasing it to drop the object. Gravity and collision allowed objects to fall above the T1 box, and both hands could carry two objects at once (Figure 5). In the experimental condition (EC), participants used Hands-Free VR interface (Figure 6), pressing the trigger button to issue voice commands and releasing it to confirm. In both conditions, participants selected objects before placing them in the box (T1) or on the floor (T2).

**Data collection:** We collected data to assess Hands-Free VR's robustness and compare it to the conventional VR interface using objective and subjective metrics. Objective metrics included task completion time (seconds), cumulative viewpoint translation (meters), view direction rotation (degrees), and controller translations. Subjective data came from a user preference questionnaire with five questions:

Q1 The interface is tedious. It requires a lot of work.
Q2 The interface is intuitive. I quickly figured out how to use it.
Q3 The interface requires a lot of physical effort.
Q4 The interface is unreliable, it often does the wrong thing.
Q5 I would love to have a similar interface on my computer.

Responses were on a five-point Likert scale, scored from 1 to 5. For negative questions (Q1, Q3, Q4), scores were reversed ($x$ replaced with $6 - x$) so higher scores always indicated better outcomes. **Research hypotheses:** We hypothesized that Hands-Free VR is robust, more efficient than a conventional VR interface, and preferred by users.

| Metric | | Time [s] | | | Viewpoint Translation [m] | | | View Rotation [deg] | | | Left Hand Translation [m] | | | Right Hand Translation [m] | | |
|---|---|---|---|---|---|---|---|---|---|---|---|---|---|---|---|---|
| Task | | T1 | T2 | T12 | T1 | T2 | T12 | T1 | T2 | T12 | T1 | T2 | T12 | T1 | T2 | T12 |
| Mean | CC | 55.44 | 64.47 | 59.95 | 18.56 | 33.67 | 26.12 | 1,385 | 1,740 | 1,562 | 12.75 | 14.89 | 13.82 | 12.67 | 13.60 | 13.14 |
| | EC | 23.76 | 22.16 | 22.96 | 0.91 | 1.04 | 0.97 | 224 | 294 | 259 | 1.16 | 1.44 | 1.30 | 0.99 | 1.07 | 1.03 |
| Std.dev. | CC | 27.56 | 14.63 | 16.93 | 6.07 | 9.57 | 6.57 | 584 | 399 | 401 | 4.86 | 3.83 | 3.67 | 4.81 | 3.49 | 3.20 |
| | EC | 14.69 | 14.90 | 11.79 | 0.78 | 0.82 | 0.64 | 153 | 227 | 156 | 0.87 | 1.29 | 0.84 | 0.84 | 0.79 | 0.69 |
| Wilcoxon | Z | -3.95 | -4.01 | -4.01 | -4.01 | -4.01 | -4.01 | -4.01 | -4.01 | -4.01 | -4.01 | -4.01 | -4.01 | -4.01 | -4.01 | -4.01 |
| | p | 0.00008 | <0.00001 | <0.00001 | <0.00001 | <0.00001 | <0.00001 | <0.00001 | <0.00001 | <0.00001 | <0.00001 | <0.00001 | <0.00001 | <0.00001 | <0.00001 | <0.00001 |
| Cohen's d | | 1.43 | 2.87 | 2.54 | 4.08 | 4.80 | 5.39 | 2.72 | 4.46 | 4.28 | 3.32 | 4.71 | 4.70 | 3.38 | 4.95 | 5.23 |

Table 1: Descriptive and inference statistics for the five objective metrics, considering the two tasks separately (T1 and T2), and together (T12), and for the conventional (CC) and Hands-Free VR (EC) conditions. In all instances, EC has a significant efficiency advantage over CC ($p < 0.05$).

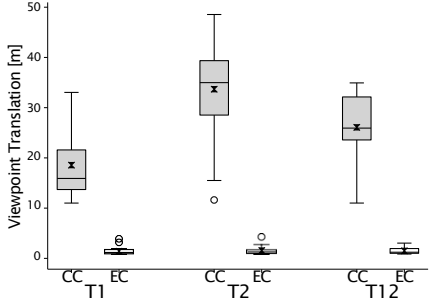

Figure 7: Viewpoint translation comparison.

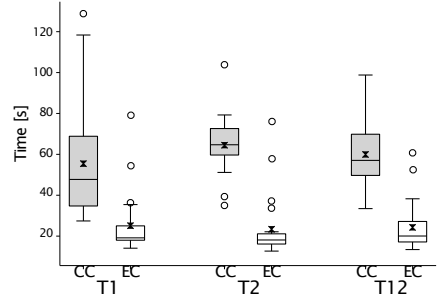

Figure 8: Task completion time between conditions.

RH1: Hands-Free VR is robust, with an overall spoken command success rate $s_{STC}$ of over 90%, i.e., $s_{STC} > 0.90$.

RH2: Participants complete both tasks faster, with less viewpoint translation, with less view direction rotation, and with less hand translation when using the Hands-Free VR interface than when using the conventional VR interface.

RH3: Participants find that the Hands-Free VR interface is less tedious and requires less physical effort than the conventional VR interface and that Hands-Free VR is intuitive and reliable.

**Procedure:** Participants completed a demographics questionnaire, practiced each task in both conditions, and then performed three trials per task in both conditions, in a counterbalanced order. They completed a preference questionnaire via VR headset after each condition. The 30-minute experiment concluded with participants receiving a USD 20 gift card.

**Data analysis:** We analyzed the data using descriptive (tables, box plots) and inferential statistics with appropriate tests with SPSS [IBM Corp., 2022]. Normality was checked via the Shapiro-Wilk test [Shapiro and Wilk, 1965]. Depending on normality, we used either the dependent t-test or the non-parametric Wilcoxon signed-rank test [Wilcoxon, 1992], suitable for our paired sample design. The Wilcoxon test handled continuous (objective metrics) and ordinal data (Likert scale responses). We set significance at $\alpha = 0.05$ and calculated effect sizes using Cohen's d [Cohen, 2013] to assess statistical power.

## 6.2 Results and Discussion

**Hands-Free VR robustness:** We measured the robustness of Hands-Free VR over all 359 verbal commands issued by our study participants. The speech-to-text conversion success rate $s_{STT} = 96.71 \pm 0.05\%$, the text-to-command success rate $s_{TTC} = 97.83 \pm 0.07\%$, for an overall verbal command success rate of $s_{STC} = s_{STT} \times s_{TTC} = 94.61\%$. This supports RH1, i.e., Hands-Free VR is robust, including with the spoken English accents of our participants, 20 of whom were not native English speakers.

**Hands-Free VR vs. conventional interface:** The interface efficiency measurements according to *the*

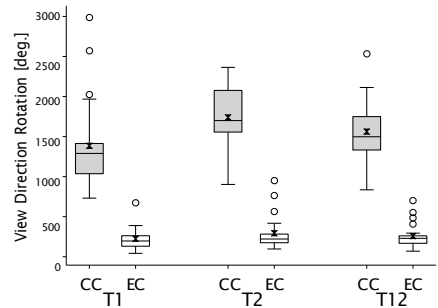

Figure 9: View direction rotation between conditions.

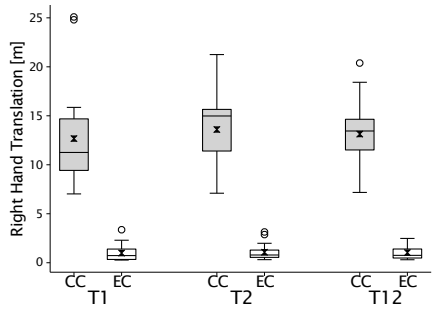

Figure 10: Right-hand translation between conditions.

*five objective metrics* are presented numerically in Table 1, and graphically, using box plots, in Figures 7 to 11.

In the box plots, the black line represents the median, the cross the mean, the bar spans the interquartile range ($q_1$ to $q_3$), whiskers show the range ($q_1 \pm 1.5 \times (q_3 - q_1)$), and dots mark outliers outside this range. Hands-Free VR enabled faster task completion with less walking, head rotation, and hand movement. The advantage was significant for T1 and even greater for T2, as the conventional VR interface required precise object placement for T2, while Hands-Free VR made both tasks equally simple with verbal commands.

With Hands-Free VR, each task required two verbal commands: one for selection and one for placement. Commands were executed in 1.51 seconds on average: $0.99 \pm 0.009$s for speech-to-text and $0.51 \pm 0.017$s for text-to-command conversion—much faster than Quest's built-in STT (2.29s per API call) Task completion time (e.g., 22 s for T2) was mostly spent reading task descriptions, and users had an improvement of 15% over each iteration of the task. In contrast, the conventional VR interface showed a linear dependency between completion time and the number of objects manipulated, despite using both hands simultaneously. It also showed a better rate of improvement than the verbal interface, at 20%. This can be explained by the user's frustration with the conventional interface, leading to rushing the task. Meanwhile, Hands-Free VR's ability to handle multiple objects in parallel gives it a growing advantage as the object count increases.

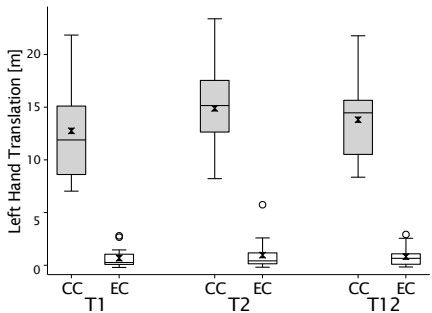

Figure 11: Left-hand translation between conditions.

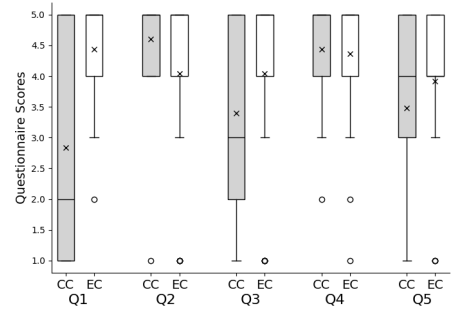

Figure 12: User preference questionnaire.

The data does not have a normal distribution, so we compared the means using Wilcoxon's signed rank test. The comparison confirms the statistical significance of the efficiency advantage of Hands-Free VR over the conventional VR user interface for each task, for both tasks combined, and for each of the five objective metrics. This provides support for RH2, i.e., Hands-Free VR has a significant efficiency advantage over the conventional VR user interface. The effect sizes measured using Cohen's $d$ are all larger than 1.0, confirming that the study's N = 22 participants are sufficient for a power of 90.

The *user preference questionnaire* results are presented in Table 2 (numerical) and Figure 12 (box plots). Negatively phrased questions were reversed ($6 - x$), so higher scores indicate better outcomes. The largest and only significant difference was for Q1, where participants found the conventional VR interface more "tedious" and requiring "more work" than Hands-Free VR. For Q3, participants strongly disagreed that Hands-Free VR required significant physical effort (median 5), while the conventional interface had a median of 3, though this difference was not significant. Despite this, participants often treated the conventional interface as a physical exercise challenge. Questions 2, 4, and 5 had high, similar median scores for both conditions (5, 5, and 4, respectively). Importantly, Hands-Free VR scored the maximum median for both intuitiveness (Q2) and reliability (Q4). We conclude that RH3 is supported, except the perceived reduction in physical effort with Hands-

| Question | | Q1 | Q2 | Q3 | Q4 | Q5 |
|---|---|---|---|---|---|---|
| **Median** | CC | 2 | 5 | 3 | 5 | 4 |
| | EC | 5 | 5 | 5 | 5 | 4 |
| **Mean** | CC | 2.84 | 4.60 | 3.4 | 4.44 | 3.48 |
| | EC | 4.44 | 4.04 | 4.04 | 4.36 | 3.92 |
| **Std. dev.** | CC | 1.68 | 0.87 | 1.47 | 0.87 | 1.39 |
| | EC | 0.82 | 1.46 | 1.62 | 1.11 | 1.26 |
| **Wilcoxon** | $Z$ | -3.57 | -1.64 | -1.49 | -0.35 | -1.53 |
| | $p$ | 0.00036 | 0.10 | 0.134 | ~1 | 0.123 |

Table 2: User preference questionnaire statistics.

Free VR was not significant.

**Discussion:** Hands-Free VR reliably interprets spoken commands, even with pauses or incomplete sentences, using a trigger button and VR-optimized models. Tested with mostly non-native English speakers, it excelled in accent robustness, outperforming Quest 3's speech-to-text, which had a Word Error Rate of 43% in our tests, by supporting diverse accents from day one. Most of our 22 participants were new to VR, probably contributing to their preference for physical interaction and viewing voice commands as efficient but less engaging.

In work settings, users may prefer voice-based interfaces for complex tasks such as handling multiple objects, precise configurations, or information retrieval, while traditional controls suit simpler tasks. For applications focused on physical exercise or skill building, such as engine assembly, core actions must remain physical. Voice commands can be supplemented by identifying parts, showing assembly order, or setting conditions, preserving embodied learning while reducing menu navigation and cognitive load.

# 7 CONCLUSIONS, LIMITATIONS, AND FUTURE WORK

We introduced Hands-Free VR, a voice-based VR interface that converts speech into executable commands. Fine-tuned for phonetic similarity, diverse accents, and natural language variation, Hands-Free VR achieved a 94% command understanding, enabling efficient execution of complex tasks and offering significant advantages over traditional VR controls.

## 7.1 Limitations

One limitation is our reliance on a powerful external workstation as an edge server. Although standalone VR headsets have advanced considerably, running speech-to-text and large language models directly on such devices is still not feasible. Our tests showed that speech-to-text took about 3 seconds and text-to-command took over 21 seconds on a laptop-class GPU (Nvidia RTX 3070), which is too slow for most VR applications. Until new technology emerges, robust voice-based VR interfaces will likely depend on a distributed server-client setup.

A limitation is the slow response time, even with a powerful edge server, which can be problematic for time-sensitive commands such as stopping a dynamic VE. Rewinding to when the user started speaking could help. Future work could speed up inference using quantization or simpler LLMs, and combining speech-to-text and text-to-command into one model could streamline processing. Another limitation is that complex commands must currently be split into simpler steps, e.g., "Select blue cubes" and "Put them into the box" instead of "Select blue cubes and move them into the box". Furthermore, the system had a limited number of discrete parameters per command, which means that continuous commands such as move x distance were not supported by our system due to the LLM's tendency to hallucinate compromising the robustness of the system.

## 7.2 Future work

Hands-Free VR can be expanded simply by adding new verbs, objects, and attributes to its syntax and lexicon, then re-running data synthesis and fine-tuning. The VR application would need the corresponding execution capabilities. Future work aims to remove the programmer from this process, letting users define and extend the interface by demonstrating desired actions. The key advantage: Users providing only one English formulation while Hands-Free VR generates multiple variants should be preserved.

A more ambitious goal is to eliminate the fixed set of commands. Instead of mapping speech to predefined commands, Hands-Free VR could generate VR interface source code on the fly. This would support an unbounded range of commands, although ensuring robustness remains challenging.

Another direction is to determine which tasks are best served by voice commands. Although our experiments highlight clear advantages for certain actions, understanding which commands are general and which are domain-specific would simplify designing future VR interfaces.

Moreover, this type of interface could be applied to systems in areas such as robotics and healthcare, where it would allow users with mobility limitations to interact with an agent or system.

An interesting direction would be the efficiency and user preference of a hybrid system, where they can experience the best of both worlds, the efficiency of voice commands with the immersion and interactivity of conventional VR interfaces.

Hands-Free VR improves VR efficiency by minimizing head movements, reducing rotations from five full turns to less than one, potentially mitigating cyber-sickness—a benefit for future study. It also supports diverse accents but should be tested with broader participant variability. Voice input adds a valuable channel, simplifying tedious tasks and their specification.

## ACKNOWLEDGEMENTS

This work was supported in part by the National Science Foundation grants 2417510 and 2309564.